\documentclass[twocolumn,aps,pra,groupedaddress,showpacs]{revtex4}
\usepackage{epsfig,amssymb,amsmath,amsthm}

\newtheorem{thm}{Theorem}

\begin{document}

\title{Classical capacity of bosonic broadcast communication \\ and a new minimum output entropy conjecture}
\author{Saikat Guha, Jeffrey H. Shapiro, and Baris I. Erkmen}\affiliation{Massachusetts Institute of Technology, Research Laboratory of
Electronics, Cambridge, Massachusetts 02139 USA}

\begin{abstract}
Previous work on the classical information capacities of bosonic channels has established the capacity of the single-user pure-loss channel, bounded the capacity of the single-user thermal-noise channel, and bounded the capacity region of the multiple-access channel.  The latter is a multi-user scenario in which several transmitters seek to simultaneously and independently communicate to a single receiver.  
We study the capacity region of the bosonic broadcast channel, in which a single transmitter seeks to simultaneously and independently communicate to two different receivers.  
It is known that the tightest available lower bound on the capacity of the single-user thermal-noise channel is that channel's capacity if, as conjectured, the minimum von Neumann entropy at the output of a bosonic channel with additive thermal noise occurs for coherent-state inputs. Evidence in support of this minimum output entropy conjecture has been accumulated, but a rigorous proof has not been obtained.  In this paper, we propose a new minimum output entropy conjecture that, if proved to be correct, will establish that the capacity region of the bosonic broadcast channel equals the inner bound achieved using a coherent-state encoding and optimum detection.   We provide some evidence that supports this new conjecture, but again a full proof is not available.  
\end{abstract}


\maketitle

\section{Introduction}

The past decade has seen several advances in evaluating classical information capacities of several important quantum communication channels \cite{bennettshor}--\cite{holevobook}. Despite these advances \cite{bennettshor}, exact capacity results are not known for many important and practical quantum communication channels. Here we extend the line of research aimed at evaluating capacities of bosonic communication channels, which began with the capacity derivation for the input photon-number constrained lossless bosonic channel \cite{yuenozawa},\cite{caves}.   The capacity of the lossy bosonic channel was found in \cite{ultcap},  where it was shown that a modulation scheme using classical light (coherent states) suffices to achieve ultimate communication rates over this channel. Subsequent attempts to evaluate the capacity of the lossy bosonic channel with additive Gaussian noise \cite{holevobook} led to a crucial conjecture on the minimum output entropy of a class of bosonic channels \cite{gglms}. Proving that conjecture would complete the capacity proof for the bosonic channel with additive Gaussian noise, and it would show that this channel's capacity is achievable with classical-light modulation.  More recent work that addressed bosonic multiple-access communication channels \cite{byen} revealed that modulation of information using non-classical states of light is necessary to achieve ultimate single-user rates in the multiple-access scenario. 

In the present work, we study the classical information capacity of the bosonic broadcast channel. A broadcast channel is the congregation of communication media connecting a single transmitter to two or more receivers. In general, the transmitter encodes and sends out independent information to each receiver in a way that each receiver can reliably decode its respective information.  We will show that when coherent-state encoding is employed in conjunction with coherent detection, the bosonic broadcast channel is equivalent to a classical degraded Gaussian broadcast channel whose capacity region is known, and known to be dual to that of the classical Gaussian multiple-access channel \cite{goldsmith}.  Thus, under these coding and detection assumptions, the capacity region of the bosonic broadcast channel is dual to that of the multiple-access bosonic channel with coherent-state encoding and coherent detection.  To treat more general transmitter and receiver conditions, we use a limiting argument to apply the degraded quantum broadcast-channel coding theorem for finite-dimensional state spaces \cite{yard} to the infinite-dimensional bosonic channel with an average photon-number constraint.  We consider the two-receiver case in which Alice ($A$) simultaneously transmits to Bob ($B$), via the transmissivity $\eta > 1/2$ port of a lossless beam splitter, and to Charlie ($C$), via that beam splitter's reflectivity $1-\eta < 1/2$ port, using arbitrary encoding and optimum measurement with an average photon number $\bar{N}$ at the input.  Given a new conjecture about the minimum output entropy of a lossy bosonic channel, we show that the ultimate capacity region is achieved by coherent-state encoding, and is given by
\begin{equation}
R_B \le g(\eta\beta\bar{N}), {\quad}R_C \le g((1-\eta)\bar{N}) - g((1-\eta)\beta\bar{N}),
\label{eq:bosoniccapregion}
\end{equation} 
for  $0\le \beta \le 1$, where $g(x) \equiv (x+1)\log(x+1)-x\log(x)$ is the von Neumann entropy of the Bose-Einstein distribution with mean $x$.  Interestingly, this capacity region is \em not\/\rm\ dual to that of the bosonic multiple-access channel with coherent-state encoding and optimum measurement that was found in \cite{byen}. 

The remainder of this paper is organized as follows.  Section~II gives a brief introduction to the capacity region of classical broadcast channels. In Sec.~III, we describe some recent work on the capacity region of the degraded quantum broadcast channel \cite{yard}. In Sec.~IV, we introduce the noiseless bosonic broadcast channel model, and derive its capacity region subject to a new minimum output entropy conjecture. In Sec.~V we compare the capacity region obtained in Sec.~IV with the classical Gaussian broadcast channel results that apply for coherent-state encoding and coherent (homodyne or heterodyne) detection.  We also show that a recent duality result between capacity regions of classical multiple-input, multiple-output Gaussian multiple-access and broadcast channels \cite{goldsmith} does \em not\/\rm\ hold for bosonic channels with coherent-state encoding.  Discussion of bosonic-channel minimum output entropy conjectures, and evidence supporting the conjecture associated with the bosonic broadcast channel, will be given in Appendix~A.

\section{Classical Broadcast Channel}

A two-user discrete, memoryless broadcast channel is modeled by a classical probability distribution, $p_{B,C\mid A}(\beta,\gamma\mid \alpha)$, where $\alpha$, $\beta$, and $\gamma$ are drawn from Alice's input alphabet ${\cal A}$, and Bob and Charlie's output alphabets ${\cal B}$, ${\cal C}$ respectively. A broadcast channel is said to be memoryless if successive uses of the channel are independent, i.e., $p_{B^n,C^n\mid A^n}(\beta^n, \gamma^n\mid \alpha^n) = \Pi_{i=1}^np_{B,C\mid A}(\beta_i,\gamma_i\mid \alpha_i)$ is the transition distribution for $n$ channel uses.  A $((2^{nR_B},2^{nR_C}),n)$ code for a broadcast channel consists of an encoder
\begin{equation}
\alpha^n: 2^{nR_B} \times 2^{nR_C} \rightarrow {\cal A}^n,
\end{equation}
and two decoders
\begin{eqnarray}
\tilde{W}_B: {\cal B}^n \rightarrow 2^{nR_B} \\
\tilde{W}_C: {\cal C}^n \rightarrow 2^{nR_C}.
\end{eqnarray}
The probability of error $P_e^{(n)}$ is the probability that the overall decoded message does not match the transmitted message, i.e.,
\begin{equation}
P_e^{(n)} = P({\tilde W}_B(B^n) \ne W_B \mbox{ or }{\tilde W}_C(C^n) \ne W_C),
\end{equation}
where the messages, $(W_B, W_C)$, that are sent to Bob and Charlie, respectively, are assumed to be uniformly distributed over the $2^{nR_B} \times 2^{nR_C}$ possibilities. A rate pair $(R_B, R_C)$ is said to be achievable, for the broadcast channel, if there exists a sequence of $((2^{nR_B},2^{nR_C}),n)$ codes with $P_e^{(n)} \rightarrow 0$ as $n \to \infty$. The capacity region of the broadcast channel is the closure of the set of achievable rates.

Determining the capacity region of a general broadcast channel is still an open problem.  The capacity region is known, however, for degraded broadcast channels, in which one receiver (say $C$) is ``downstream'' from the first receiver (say $B$), so that $C$ always receives a degraded version of what $B$ observes. In other words, $A \rightarrow B \rightarrow C$ is a Markov chain, so that there exists a distribution, ${\tilde p}(\gamma\mid \beta)$, such that
\begin{equation}
p_{C\mid A}(\gamma \mid \alpha) = \sum_{\beta}p_{B \mid A}(\beta \mid \alpha){\tilde p}(\gamma \mid \beta).
\end{equation}
Degraded broadcast channels were first studied by Cover \cite{cover}, who conjectured that the capacity region for Alice to send independent information to Bob and Charlie at rates $R_B$ and $R_C$ respectively over such a channel is the convex hull of the closure of all $(R_B, R_C)$ satisfying 
\begin{eqnarray}
R_B &\le& I(A;B \mid T) \\
R_C &\le& I(T;C),
\label{broadcastregion}
\end{eqnarray}
for some joint distribution $p_T(t)p_{A \mid T}({\alpha} \mid t)p_{B,C \mid A}(\beta,\gamma \mid \alpha)$.  Here, $I(X; Y)$ denotes the Shannon mutual information between ensembles $X$ and $Y$, and $T$ is an auxiliary random variable with cardinality $|{\cal T}| \le \min\left\{|{\cal A}|,|{\cal B}|,|{\cal C}|\right\}$. The achievability of the above capacity result was proved by Bergmans \cite{bergmans}, and Gallager proved the converse \cite{gallager}.

\section{Quantum Degraded Broadcast Channel}

A quantum channel ${\cal N}_{A\text{-}B}$ from Alice to Bob is a trace-preserving completely positive map that transforms Alice's single-use density operator ${{\hat{\rho}}}^A$ into Bob's:  ${{\hat{\rho}}}^B = {\cal N}_{A\text{-}B}({\hat{\rho}}^A)$. The two-user quantum broadcast channel ${\cal N}_{A\text{-}BC}$ is a quantum channel from sender Alice ($A$) to two independent receivers, Bob ($B$) and Charlie ($C$). The quantum channel from Alice to Bob is obtained by tracing out $C$ from the channel map, i.e., ${\cal N}_{A\text{-}B} \equiv {\rm Tr}_C\left({\cal N}_{A\text{-}BC}\right)$, with a similar definition for ${\cal N}_{A\text{-}C}$. We say that a broadcast channel ${\cal N}_{A\text{-}BC}$ is degraded if there exists a degrading channel, ${\cal N}^{\rm {deg}}_{B\text{-}C}$, from $B$ to $C$ satisfying
${\cal N}_{A\text{-}C} = {\cal N}^{\rm {deg}}_{B\text{-}C} \circ {\cal N}_{A\text{-}B}.$
The degraded broadcast channel describes a physical scenario in which for each successive $n$ uses of ${\cal N}_{A\text{-}BC}$ Alice communicates a randomly generated classical message $(m,k) \in (W_B, W_C)$ to Bob and Charlie, where the message sets $W_B$ and $W_C$ have cardinalities $2^{nR_B}$ and $2^{nR_C}$ respectively. The messages $(m,k)$ are assumed to be uniformly distributed over $(W_B, W_C)$.  Because of the degraded nature of the channel, Bob receives both $m$ and $k$, whereas Charlie only receives $k$. 

To convey the message $(m,k)$, Alice prepares an $n$-channel-use input state, with density operator $\hat{\rho}_{m,k}^{A^n}$, from ${\cal{A}}^n$, the tensor product space of her single-use input-state alphabet.  After transmission through the channel, this state results in the bipartite density operator $\hat {\rho}_{m,k}^{B^nC^n} = {\cal{N}}_{A\text{-}BC}^{\otimes n}(\hat{\rho}_{m,k}^{A^n})$ for Bob and Charlie.  The reduced density operators for Bob and Charlie, $\hat{\rho}_{m,k}^{B^n}$ and $\hat{\rho}_{m,k}^{C^n}$ respectively, can be found by tracing out the other receiver.  A $(2^{nR_B}, 2^{nR_C}, n, \epsilon)$ code for this channel consists of an encoder
\begin{equation}
(m,k): (W_B,W_C) \rightarrow {\cal A}^n,
\label{eq:encoder}
\end{equation}
a positive operator-valued measure (POVM) $\left\{\Lambda_{mk}\right\}$ on ${\cal B}^n$ and a POVM $\left\{\Lambda_{k}^\prime\right\}$ on ${\cal C}^n$ that satisfy \cite{footnote1}
\begin{equation}
{\rm Tr}\left(\hat{\rho}_{m,k}^{B^nC^n}(\Lambda_{mk} \otimes \Lambda_{k}^\prime)\right) \ge 1-\epsilon,
\label{eq:dec_condition}
\end{equation}
for all $(m,k) \in (W_B,W_C)$.  Its error probability therefore obeys $P_e^{(n)} \le \epsilon$. A rate-pair $(R_B,R_C)$ is achievable if there exists a sequence of $(2^{nR_B},2^{nR_C},n,\epsilon_n)$ codes with $\epsilon_n \rightarrow 0$, so that $P_e^{(n)} \rightarrow 0$ for such a sequence. The classical capacity region of the degraded quantum broadcast channel is then the convex hull of the closure of all achievable rate pairs $(R_B, R_C)$. 

\begin{figure}
\begin{center}
\includegraphics[width=3in]{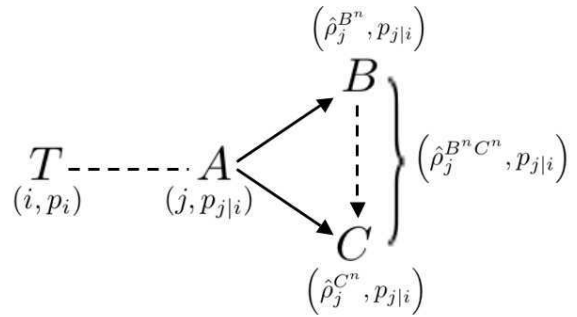}
\end{center}
\caption{Schematic of the degraded quantum broadcast channel. The transmitter Alice ($A$) encodes her messages to Bob ($B$) and Charlie ($C$) in a classical index $j$ by preparing $B$ and $C$ in the bipartite state ${\hat \rho}_j^{B^nC^n}$.  The dashed line from $B$ to $C$ denotes the existence of a degrading channel, ${\cal N}^{\rm deg}_{B\text{-}C}$, whose $n$-fold tensor product will transform  $\hat{\rho}_j^{B^n}$ into $\hat{\rho}_j^{C^n}$ for all $j$.}
\label{fig:TABC}
\end{figure}

The classical capacity region of the two-user degraded quantum broadcast channel ${\cal N}_{A\text{-}BC}$ was recently derived by Yard {\em et al.} \cite{yard}, and can be expressed in terms of the Holevo information \cite{holevo}, 
\begin{equation}
\chi\!\left({p_j, {  {\hat{\sigma}}}_j}\right) \equiv S\!\left(\sum_j{p_j{  {\hat{\sigma}}}_j}\right) - \sum_jp_jS({  {\hat{\sigma}}}_j),
\end{equation}
where $\left\{p_j\right\}$ is a probability distribution associated with the density operators $\{  {\hat{\sigma}}_j\}$, and $S({\hat{\rho}}) \equiv -{\rm {Tr}}({\hat{\rho}}\log{\hat{\rho}})$ is the von Neumann entropy of the quantum state ${\hat{\rho}}$. 
Because $\chi$ may not be additive, the rate region $(R_B, R_C)$ of the degraded broadcast channel must be computed by maximizing over multiple channel uses.  Thus, for  $n$ channel uses we can achieve the rate region specified by
\begin{eqnarray}
R_B &\le& \sum_ip_i\chi\!\left(p_{j\mid i},{\cal N}_{A\text{-}B}^{\otimes n}({\hat \rho}_j^{A^n})\right)/n \nonumber \\
&=& \frac{1}{n}\sum_ip_i\left[S\!\left(\sum_jp_{j\mid i}{\hat \rho}_j^{B^n}\right) \right. \nonumber \\
&-& \left.\sum_{j}p_{j\mid i}S\!\left({\hat \rho}_j^{B^n}\right)\right],   \label{eq:RB-bound_chi} \\
R_C &\le& \chi\!\left(p_i, \sum_jp_{j\mid i}{\cal N}_{A\text{-}C}^{\otimes n}({\hat \rho}_j^{A^n})\right)/n \nonumber \\
&=& \frac{1}{n}\left[ S\!\left(\sum_{i,j}p_ip_{j\mid i}{\hat \rho}_j^{C^n}\right) \right. \nonumber \\
&-& \left. \sum_ip_iS\!\left(\sum_jp_{j\mid i}{\hat \rho}_j^{C^n}\right) \right], \label{eq:RC-bound_chi} 
\end{eqnarray}
where $j \equiv (m,k)$ is a collective index. The probabilities $\left\{p_i\right\}$ form a distribution over an auxiliary classical alphabet ${\cal T}$, of size $|{\cal T}|$, satisfying $|{\cal T}| \le \min\left\{|{\cal A}^n|,|{\cal B}^n|^{2} + |{\cal C}^n|^{2} + 1\right\}$. The ultimate rate-region is computed by maximizing the region specified by Eqs.~\eqref{eq:RB-bound_chi} and \eqref{eq:RC-bound_chi}, over $\left\{p_i\right\}$, $\left\{p_{j|i}\right\}$, $\left\{{\hat \rho}_j^{A^n}\right\}$, and $n$, subject to the cardinality constraint on $|{\cal T}|$. Figure~\ref{fig:TABC} illustrates the setup of the two-user degraded quantum channel.

\section{Noiseless bosonic Broadcast Channel}

The two-user noiseless bosonic broadcast channel ${\cal N}_{A\text{-}BC}$ consists of a collection of spatial and temporal bosonic modes at the transmitter (Alice) that interact with a minimal-quantum-noise environment and split into two sets of spatio-temporal modes en route to two independent receivers (Bob and Charlie). The multi-mode two-user bosonic broadcast channel ${\cal N}_{A\text{-}BC}$ is given by $\bigotimes_s{{\cal N}_{A_s\text{-}B_sC_s}}$, where ${{\cal N}_{A_s\text{-}B_sC_s}}$ is the broadcast-channel map for the $s$th mode, which can be obtained from the Heisenberg evolutions
\begin{eqnarray}
{\hat b}_s &=& {\sqrt {\eta_s}}\,{\hat a}_s + {\sqrt {1-\eta_s}}\,{\hat e}_s, \label{eq:BS-modebk} \\[.05in]
{\hat c}_s &=& {\sqrt {1-\eta_s}}\,{\hat a}_s - {\sqrt {\eta_s}}\,{\hat e}_s, \label{eq:BS-modeck}
\end{eqnarray}
where $\{{\hat a}_s\}$ are Alice's modal annihilation operators, and $\{{\hat b}_s\}$, $\{{\hat c}_s\}$ are the corresponding modal annihilation operators for Bob and Charlie, respectively. The modal transmissivities $\{\eta_s\}$ satisfy $0 \le \eta_s \le 1$, and the environment modes $\{{\hat e}_s\}$ are in their vacuum states. We will limit our treatment here to the single-mode bosonic broadcast channel, as the capacity of the multi-mode channel can in principle be obtained by summing up capacities of all spatio-temporal modes and maximizing the sum capacity region subject to an overall input-power budget using Lagrange multipliers, cf.~\cite{holevobook}, where this was done for the capacity of the multi-mode single-user lossy bosonic channel.

\begin{figure}
\begin{center}
\includegraphics[width=3.in]{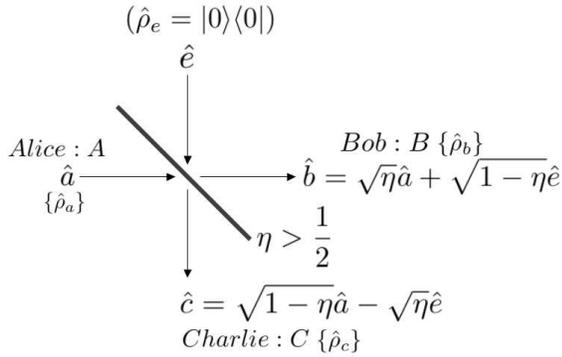}
\end{center}
\caption{(Color online) A single-mode noiseless bosonic broadcast channel can be envisioned as a beam splitter with transmissivity $\eta$. With $\eta > 1/2$, the bosonic broadcast channel is a degraded quantum broadcast channel, where Bob ($B$) is the less-noisy receiver and Charlie ($C$) is the more-noisy receiver.}
\label{fig:qbc}
\end{figure}

The principal result we have for the single-mode degraded bosonic broadcast channel depends on a minimum output entropy conjecture (strong conjecture~2, see Appendix~A).  Assuming this conjecture to be true, we will show that the ultimate capacity region of the single-mode noiseless bosonic broadcast channel (see Fig.~\ref{fig:qbc}) with a mean input photon-number constraint $\langle{\hat a}^\dagger{\hat a}\rangle \le {\bar N}$ is
\begin{eqnarray}
R_B &\le& g(\eta\beta{\bar N}),  \label{eq:RBultrate}\\[.05in]
R_C &\le& g((1-\eta){\bar N}) - g((1-\eta)\beta{\bar N}). \label{eq:RCultrate}
\end{eqnarray}
Here, $0 \le \beta \le 1$ is a parameter that represents the fraction of Alice's average photon number that is used to convey information to Bob, with remainder to be used to communicate information to Charlie.  The boundary of the broadcast channel's capacity region is traced out by varying $\beta$ from 0 to 1.  

It is worth noting, at this point, that our assumption of a lossless beam splitter---in Eqs.~\eqref{eq:BS-modebk} \eqref{eq:BS-modeck}, and Fig.~\ref{fig:qbc}---is not essential.  In particular, if the coupling coefficients from  $\hat{a}$ to $\hat{b}$ and $\hat{a}$ to $\hat{c}$ in Fig.~\ref{fig:qbc} were $\eta_b$ and $\eta_c$, respectively, with $0 \le \eta_c <  \eta_b$, and $\eta_b + \eta_c < 1$, then we still have a degraded quantum broadcast channel, and, assuming strong conjecture~2 is correct, its capacity region is given by Eqs.~\eqref{eq:RBultrate} and \eqref{eq:RCultrate}, with $\eta_b$ and $\eta_c$ replacing $\eta$ and $1-\eta$, respectively.   For simplicity, in all that follows, we shall presume that the lossless beam splitter model applies.  

The rate region from Eqs.~\eqref{eq:RBultrate} and \eqref{eq:RCultrate} is additive and achievable with single channel use coherent-state encoding using the distributions
\begin{eqnarray}
p_T(t) &=& \frac{1}{\pi{\bar N}}\exp\left(-\frac{|t|^2}{\bar N}\right), 
\label{opt-dist1} \\[.05in]  
p_{A\mid T}(\alpha \mid t) &=& \frac{1}{{\pi}{\bar N}\beta}\exp\left(-\frac{|{\sqrt {1-\beta}}\,t-\alpha|^2}{{\bar N}\beta}\right).
\label{opt-dist2}
\end{eqnarray}

The first step toward proving that Eqs.~\eqref{eq:RBultrate} and \eqref{eq:RCultrate} do indeed specify the bosonic broadcast channel's capacity region is to show that Eqs.~\eqref{opt-dist1} and \eqref{opt-dist2} achieve these rates.  It is straightforward to show that if $\eta > 1/2$, the bosonic broadcast channel is a degraded quantum broadcast channel, in which Bob's is the less-noisy receiver and Charlie's is the more-noisy receiver. To do so we merely recognize that, when $\eta > 1/2$, Charlie's reduced density operator can be obtained from Bob's by applying $\{\hat{b}_i : 1\le i \le n\}$ to the input of a lossless beam splitter that has transmissivity $\eta' = (1-\eta)/\eta$ to output modes $\{\hat{c}_i : 1\le i \le n\}$, and whose other input port is driven by vacuum-state modes.  The Yard {\em et al.} capacity region in Eqs.~\eqref{eq:RB-bound_chi} and \eqref{eq:RC-bound_chi} requires finite-dimensional Hilbert spaces for the channel's input and outputs. Nevertheless, we will use their result for the bosonic broadcast channel, which has an infinite-dimensional state space, by extending it to infinite-dimensional state spaces through a limiting argument \cite{footnote2}. The $n$ = 1 rate-region for the bosonic broadcast channel using a coherent-state encoding is thus:
\begin{eqnarray}
R_B &\le& \int{p_T(t)S\!\left(\int{p_{A\mid T}(\alpha \mid t)}|{\sqrt \eta}\,\alpha\rangle\langle{\sqrt \eta}\,\alpha|\,d\alpha\right)dt} \label{coh-capacity1} \\
R_C &\le& S\!\left(\int{p_T(t)p_{A \mid T}(\alpha\mid t)|{\sqrt {1-\eta}}\,\alpha\rangle\langle{\sqrt {1-\eta}}\,\alpha|\,d\alpha\,{dt}}\right) \nonumber \\ 
&-&{\int}p_T(t)S\!\left({\int}p_{A\mid T}({\alpha}\mid t) \right.  \nonumber \\
&\times& \left. |{\sqrt {1-\eta}}\,\alpha\rangle\langle{\sqrt {1-\eta}}\,{\alpha}|\,d\alpha\right)dt, \label{coh-capacity2}
\end{eqnarray}
where we need to maximize the bounds for $R_B$ and $R_C$ over all joint distributions $p_T(t)p_{A\mid T}(\alpha\mid t)$ subject to $\langle|\alpha|^2\rangle \le {\bar N}$. Note that $A$ and $T$ are complex-valued random variables, and the second term in the $R_B$ bound \eqref{eq:RB-bound_chi} vanishes, because the von Neumann entropy of a pure state is zero. Substituting Eqs.~\eqref{opt-dist1} and \eqref{opt-dist2} into Eqs.~\eqref{coh-capacity1} and \eqref{coh-capacity2} shows that the rate region in Eqs.~\eqref{eq:RBultrate} and \eqref{eq:RCultrate} is achievable with single-use coherent-state encoding.

For the converse, assume that the rate pair $(R_B, R_C)$ is achievable. Let $\left\{(m,k)\right\}$ and the POVMs $\left\{\Lambda_{mk}\right\}$, $\left\{\Lambda_k^\prime\right\}$ comprise any $(2^{nR_B}, 2^{nR_C}, n, \epsilon_n)$ code in the achieving sequence. Suppose that Bob and Charlie store their decoded messages in the classical registers ${\tilde W}_B$ and ${\tilde W}_C$ respectively. Let us use $p_{W_B,W_C}(m,k) = p_{W_B}(m)p_{W_C}(k) = 2^{-nR_B}2^{-nR_C}$ to denote the joint probability mass function of the independent message registers $W_B$ and $W_C$.  As $(R_B,R_C)$ is an achievable rate-pair, there must exist $\epsilon_n^\prime \to 0$, such that
\begin{eqnarray}
nR_C &=& H(W_C) \nonumber \\
&\le& I(W_C; {\tilde W}_C) + n\epsilon_n^\prime \nonumber \\
&\le& \chi(p_{W_C}(k),{\hat \rho}_k^{C^n}) + n\epsilon_n^\prime,
\end{eqnarray}
where $I(W_C; {\tilde W}_C) \equiv H({\tilde W}_C) - H({\tilde W}_C\mid W_C)$ gives the Shannon mutual information in terms of the Shannon entropy $H = -\sum_k p_k\log(p_k)$ for a probability distribution $\{p_k\}$, and $\hat{\rho}^{C^n}_k = \sum_m p_{W_B}(m)\hat{\rho}_{m,k}^{C^n}$. The second line follows from Fano's inequality \cite{Fanoineq} and the third line follows from Holevo's bound \cite{footnote3}. Similarly, for $\epsilon_n^{\prime\prime} \to 0$, we can bound $nR_B$ as follows:
\begin{eqnarray}
nR_B &=& H(W_B) \nonumber \\
&\le& I(W_B;{\tilde W}_B) + n\epsilon_n^{\prime\prime}  \nonumber \\
&\le& \chi(p_{W_B}(m),{\hat \rho}_m^{B^n}) + n\epsilon_n^{\prime\prime}  \nonumber \\
&\le& \sum_kp_{W_C}(k)\chi(p_{W_B}(m),{\hat \rho}_{m,k}^{B^n}) + n\epsilon_n^{\prime\prime},
\end{eqnarray}
where the three lines above follow from Fano's inequality, Holevo's bound and the concavity of Holevo information. 

To complete the converse proof, we need only show that there exists a $0 \le \beta \le 1$, such that 
\begin{eqnarray}
\sum_kp_{W_C}(k)\chi(p_{W_B}(m),{\hat \rho}_{m,k}^{B^n}) \le ng(\eta\beta{\bar N}), \\
\chi(p_{W_C}(k),{\hat \rho}_k^{C^n}) \le ng((1-\eta){\bar N}) - ng((1-\eta)\beta{\bar N}). 
\end{eqnarray}
From the non-negativity of the von Neumann entropy $S\!\left({\hat \rho}_{m,k}^{B^n}\right)$, it follows that $\sum_kp_{W_C}(k)\chi(p_{W_B}(m),{\hat \rho}_{m,k}^{B^n}) \le\sum_kp_{W_C}(k)S\!\left(\sum_mp_{W_B}(m){\hat \rho}_{m,k}^{B^n}\right)$, as the second term of the Holevo information above is non-negative. Because von Neumann entropy is subadditive, and the maximum von Neumann entropy of a single-mode bosonic state with $\langle{\hat a}^\dagger{\hat a}\rangle \le {\bar N}$ is given by $g({\bar N})$, we have that
\begin{equation}
0 \le S\!\left({\hat \rho}_k^{B^n}\right) \le \sum_{\ell=1}^ng\!\left(\eta{\bar N}_{k_\ell}\right) \le ng\!\left(\eta{\bar N}_k\right),
\end{equation}
where, ${\bar N}_k \equiv \sum_{\ell =1}^n{\bar N}_{k_\ell}/n$, and ${\bar N}_{k_\ell}$ is the mean photon number of the $\ell$th channel use for the state ${\hat \rho}_k^{B^n} = \sum_m p_{W_B}(m)\hat{\rho}_{m,k}^{B^n}$.  Therefore, because $g(0) = 0$ and $g(x)$ is monotonically increasing for $x > 0$, we see that for each $k \in W_C$ there is a $0\le \beta_k \le 1$ such that
\begin{equation}
S\left({\hat \rho}_k^{B^n}\right) = ng\left(\eta{\beta_k}{\bar N}_k\right).
\label{eq:eta-betak}
\end{equation}
We know that $\bar N$ is Alice's maximum average photon number per channel use, where the averaging is over the entire codebook. Thus, the mean photon number of the $n$-use average codeword at Bob, ${\hat \rho}^{B^n} \equiv \sum_kp_{W_C}(k){\hat \rho}_k^{B^n}$, is $\eta{\bar N}$. Hence, we get
\begin{equation}
0 \le \sum_kp_{W_C}(k)S\!\left({\hat \rho}_k^{B^n}\right) \le S({\hat \rho}^{B^n}) \le ng\left(\eta{\bar N}\right),
\end{equation}
where the second inequality follows from the convexity of von Neumann entropy.  Again invoking the monotonicity of $g(x)$ we have that there is a $0 \le \beta \le 1$, such that $\sum_kp_{W_C}(k)S\!\left({\hat \rho}_k^{B^n}\right) = ng(\eta\beta{\bar N})$ whence
\begin{equation}
\sum_kp_{W_C}(k)\chi(p_{W_B}(m),{\hat \rho}_{m,k}^{B^n}) \le ng(\eta\beta{\bar N}).
\label{eq:RBbeta}
\end{equation}
This proves the first inequality that we need for the capacity region's converse statement.  
  
To prove the second inequality needed for that converse, we start from Eq.~\eqref{eq:eta-betak}
and use strong conjecture~2 (see Appendix~A) to get
\begin{equation}
S\!\left({\hat \rho}_k^{C^n}\right) \ge ng\!\left((1-\eta){\beta_k}{\bar N}_k\right).
\label{eq:one-m-eta-betak}
\end{equation}
Next, we use the uniform distribution $p_{W_C}(k) = 2^{-nR_C}$ to obtain
\begin{equation}
\sum_k 2^{-nR_C}g\left(\eta\beta_k{\bar N}_k\right) = g\left(\eta\beta{\bar N}\right).
\label{eq:thm1}
\end{equation}
Using \eqref{eq:thm1}, the convexity of $g(x)$ and $\eta > 1/2$, we have shown (see Appendix~B) that
\begin{equation}
\sum_k 2^{-nR_C}g\left((1-\eta)\beta_k{\bar N}_k\right) \ge g\left((1-\eta)\beta{\bar N}\right).
\label{eq:thm2}
\end{equation}
From Eq.~\eqref{eq:thm2}, and Eq.~\eqref{eq:one-m-eta-betak} summed over $k$, we then obtain
\begin{equation}
\sum_kp_{W_C}(k)S\left({\hat \rho}_k^{C^n}\right) \ge ng((1-\eta)\beta{\bar N}).
\label{eq:minoutent1}
\end{equation}
Finally, writing Charlie's Holevo information as
\begin{eqnarray}
\lefteqn{\chi(p_{W_C}(k),{\hat \rho}_k^{C^n}) = }
\nonumber \\
&&S\!\left(\sum_kp_{W_C}(k){\hat \rho}_k^{C^n}\right) - \sum_kp_{W_C}(k)S\!\left({\hat \rho}_k^{C^n}\right) \nonumber \\
&\le& ng((1-\eta){\bar N}) 
- \sum_kp_{W_C}(k)S\!\left({\hat \rho}_k^{C^n}\right),  \label{eq:minoutent2}
\end{eqnarray}
we can use Eq.~\eqref{eq:minoutent1} to get
\begin{equation}
\chi(p_{W_C}(k),{\hat \rho}_k^{C^n}) \le ng((1-\eta){\bar N}) - ng((1-\eta)\beta{\bar N}),
\label{eq:RCbeta}
\end{equation}
which completes  proof of the converse, given that strong conjecture~2 is true.

\section{Discussion}

\begin{figure}
\begin{center}
\includegraphics[width=2.75in]{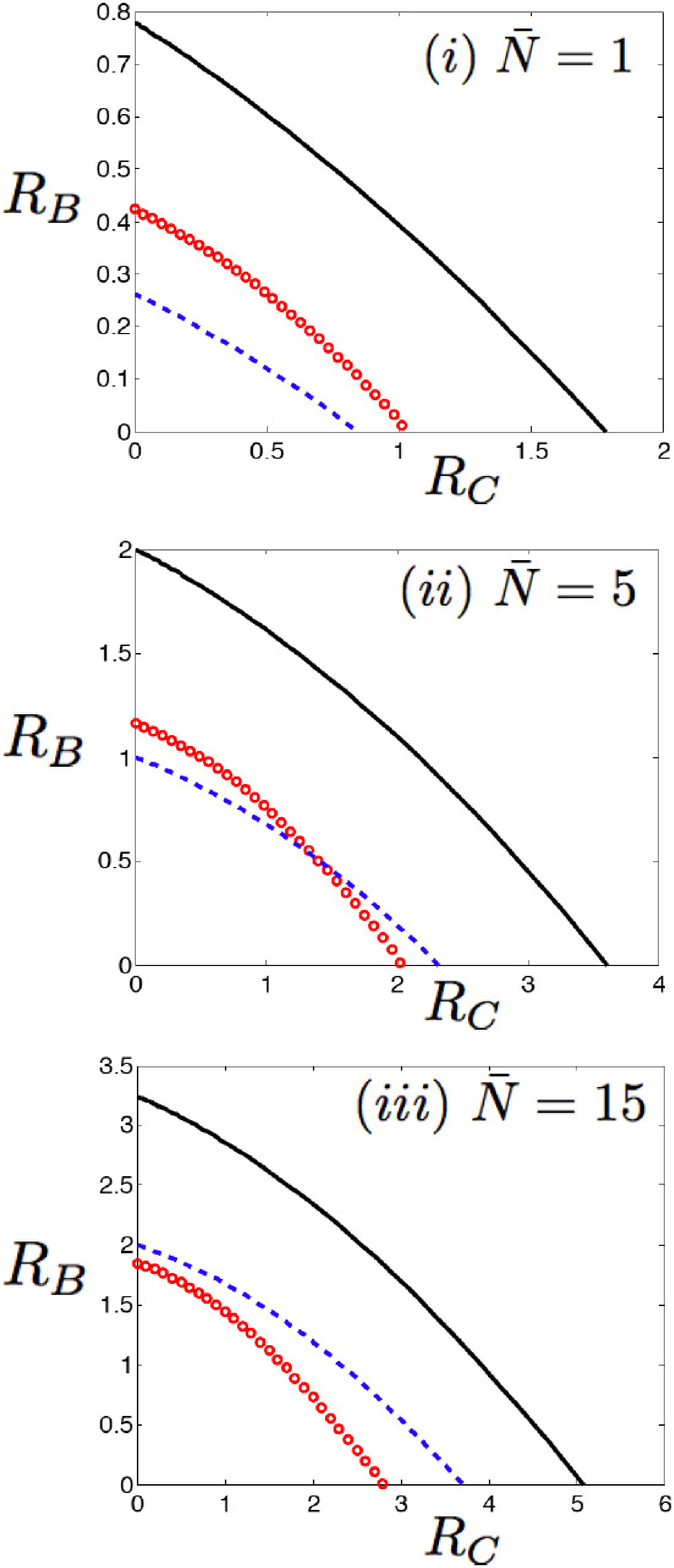}
\end{center}
\caption{(Color online) Comparison of bosonic broadcast channel capacity regions, in bits per channel use, achieved by coherent-state encoding with homodyne detection (the capacity region lies inside the boundary marked by circles), heterodyne detection (the capacity region lies inside the boundary marked by dashes), and optimum reception (the capacity region lies inside the region bounded by the solid curve), for ${\eta = 0.8}$, and ${\bar N} = 1$, $5$, and $15$.}
\label{fig:capregionshomhet}
\end{figure}

Without a proof of strong conjecture~2, we cannot assert that Eqs.~\eqref{eq:RBultrate} and \eqref{eq:RCultrate} define the capacity region of the bosonic broadcast channel.  However, because the rate region specified by these equations is achievable---with single-use coherent-state encoding---we know that they comprise an inner bound on that capacity region.  In this regard it is instructive to examine how the rate region defined by Eqs.~\eqref{eq:RBultrate} and \eqref{eq:RCultrate} compares with what can be realized by conventional, coherent-detection optical communications.  Suppose Alice sends a coherent state, $|\alpha\rangle$, to the beam splitter shown in Fig.~\ref{fig:qbc}.  Bob and Charlie will then receive coherent states $|\sqrt{\eta}\,\alpha\rangle$ and $|\sqrt{1-\eta}\,\alpha\rangle$, respectively.  
Moreover, if Bob and Charlie employ homodyne-detection receivers, with local oscillator phases set to observe the real-part quadrature, their post-measurement data will be $\sqrt{\eta}\,{\rm Re}(\alpha) + v_B$ for Bob and $\sqrt{1-\eta}\,{\rm Re}(\alpha) + v_C$ for Charlie, where $v_B$ and $v_C$ are independent, identically distributed, real-valued Gaussian random variables that are zero mean and have variance 1/4 \cite{PartIII}.  Similarly, if Bob and Charlie employ heterodyne-detection receivers, their post-measurement data will be $\sqrt{\eta}\,\alpha +  z_B$ and $\sqrt{1-\eta}\,\alpha + z_C$,  where $z_B$ and $z_C$ are independent, identically-distributed, complex-valued Gaussian random variables that are zero mean, isotropic, and have variance 1/2 \cite{PartIII}.  These results imply that the $\eta > 1/2$ bosonic broadcast channel with coherent-state encoding and homodyne detection is a classical degraded scalar-Gaussian broadcast channel, whose capacity region is known to be \cite{CoverThomas}
\begin{eqnarray}
R_B & \le & \frac{1}{2}\log(1 + 4\eta\beta \bar{N}) \\[.05in]
R_C &\le & \frac{1}{2}\log\!\left(1 + \frac{4(1-\eta)(1- \beta)\bar{N})}{1+ 4(1-\eta)\beta \bar{N}}\right),
\end{eqnarray}
for $0\le \beta \le 1$.  
Likewise, the $\eta > 1/2$ bosonic broadcast channel with coherent-state encoding and heterodyne detection is a classical degraded vector-Gaussian broadcast channel, whose capacity region is known to be 
\begin{eqnarray}
R_B & \le & \log(1 + \eta\beta \bar{N}) \\[.05in]
R_C &\le & \log\!\left(1 + \frac{(1-\eta)(1- \beta) \bar{N}}{1+ (1-\eta)\beta \bar{N}}\right),
\end{eqnarray}
for $0\le \beta \le 1$.  
In Fig.~\ref{fig:capregionshomhet} we compare the capacity regions attained by a coherent-state input alphabet using homodyne detection, heterodyne detection, and optimum reception. As is known for single-user bosonic communications, homodyne detection performs better than heterodyne detection when the transmitters are starved for photons, because it has lower noise.  Conversely, heterodyne detection outperforms homodyne detection when the transmitters are photon rich, because it has a factor-of-two bandwidth advantage. To bridge the gap between the coherent-detection capacity regions and the ultimate capacity region, one must use joint detection over long codewords. Future investigation will be needed to develop receivers that can approach the ultimate communication rates over the bosonic broadcast channel. 

\begin{figure}
\begin{center}
\includegraphics[width=3.25in]{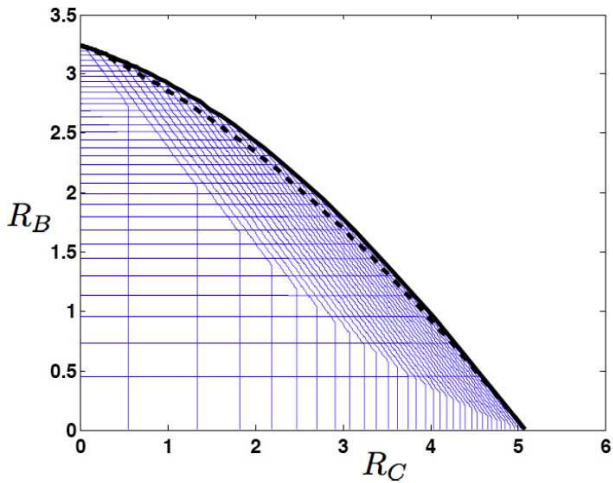}
\end{center}
\caption{(Color online) Comparison of bosonic broadcast and multiple-access channel capacity regions, in bits per channel use, for ${\eta = 0.8}$, and ${\bar N} = 15$ with coherent-state encoding.  The dashed curve is the outer boundary of the broadcast capacity region.  It lies below the solid curve, which is the outer envelope of the MAC capacity regions.}
\label{fig:dualcheck}
\end{figure}

Recently, Vishwanath {\em et al.} \cite{goldsmith} established the duality between the dirty-paper achievable rate region---recently proved to be the ultimate capacity region \cite{shamai2006}---for the classical multiple-input, multiple-output (MIMO) Gaussian broadcast channel and the capacity region of the classical MIMO Gaussian multiple-access channel (MAC).  Their duality result states that if we evaluate the capacity regions of the MIMO Gaussian MACs---with fixed total received power $P$ and channel-gain values---over all possible power allocations between the users, the corners of those capacity regions trace out the capacity region of the MIMO Gaussian broadcast channel with transmitter power $P$ and the same channel-gain values. Of course, the bosonic broadcast channel and the bosonic multiple-access channel satisfy this duality when they employ coherent-state encoding and coherent detection, because under these conditions these quantum channels reduce to classical additive Gaussian-noise channels.  However, it turns out that the capacity region of the bosonic broadcast channel using coherent-state inputs and optimum reception is \em not\/\rm\ equal to of the envelope of the MAC capacity regions using coherent-state inputs.  The capacity region of the bosonic MAC using coherent-state inputs was first computed by Yen \cite{byen}.   In Fig.~\ref{fig:dualcheck} we compare the envelope of coherent-state MAC capacities to the capacity region of the coherent-state broadcast channel.   This figure shows that with a fixed beam splitter and identical average photon number budgets, more collective classical information can be sent when the beam splitter is used as a multiple-access channel as opposed to when it is used as a broadcast channel if coherent-state encoding is employed.

\section*{Acknowledgment}

This research was supported by the Defense Advanced Research Projects Agency and by the W. M. Keck Foundation Center for Extreme Quantum Information Theory. 

\appendix
\section{Minimum Output Entropy Conjectures}

In general, the evolution of a quantum state resulting from the state's propagation through a quantum communication channel is not unitary, so that a pure state loses some coherence in its transit through the channel. The minimum von Neumann entropy at the channel's output provides a useful measure of the channel's ability to preserve the coherence of its input state.  In particular, the output entropy associated with a pure state measures the entanglement that such a state establishes with the environment during propagation through the channel. Because the state of the environment is not accessible, this entanglement is responsible for the loss of quantum coherence and hence for the injection of noise into the channel output. The study of minimum output entropy yields important information about channel capacities, viz.,  an upper bound on the classical capacity derives from a lower bound on the output entropy of multiple channel uses, see, e.g., \cite{holevobook}.  Also, the additivity of the minimum output entropy implies the additivity of the classical capacity and of the entanglement of formation \cite{shor-holevo_addivity}.

In this appendix we first briefly review previous work on a minimum output entropy conjecture that arose in conjunction with the channel capacity analysis of the single-user bosonic channel with additive Gaussian noise \cite{holevobook},\cite{gglms},\cite{renyiproof}--\cite{Eisert}.  Then we will turn our attention to the minimum output entropy conjecture that we employed in our capacity theorem for the degraded bosonic broadcast channel.  Both conjectures have  weak (single-use) and strong (multiple-use) versions.

Let $\hat{a}$ and $\hat{b}$ denote the two input modes of a lossless beam splitter of transmissivity $\eta$, that has output modes $\hat {c} = \sqrt{\eta} \,\hat{a} + \sqrt{1-\eta}\,\hat{b}$ and $\hat{d} = \sqrt{1-\eta}\,\hat{a} - \sqrt{\eta}\,\hat{b}$. In \cite{gglms}, the following minimum output entropy conjecture was made.\\ 

\noindent{\bf {Conjecture 1}} --- {\em Let the input mode $\hat{b}$ be in a thermal state with average photon number $K$ (hence von Neumann entropy $g(K)$). Then the von Neumann entropy of the output mode $\hat{c} = \sqrt{\eta}\,\hat{a} + \sqrt{1-\eta}\,\hat{b}$ is minimized when the input mode $\hat{a}$ is in the vacuum state.  The resulting minimum von Neumann output entropy is $g((1-\eta)K)$.}\\ 

The above conjecture is a special case of the following strong (multi-mode) version whose proof would establish the ultimate capacity of the single-user bosonic channel with thermal noise.\\

\noindent{\bf {Strong Conjecture 1}} --- {\em Let the input modes $\hat{\boldsymbol b} = [\begin{array}{cccc} \hat{b}_1 & \hat{b}_2 & \cdots & \hat{b}_n\end{array}]^T$ be in a product state of $n$  thermal states with total von Neumann entropy $ng(K)$. Then the von Neumann entropy of the output modes $\hat{\boldsymbol c} = [\begin{array}{cccc} \hat{c}_1 & \hat{c}_2 & \cdots & \hat{c}_n\end{array}]^T = \sqrt{\eta}\,\hat{\boldsymbol a} + \sqrt{1-\eta}\,\hat{\boldsymbol b}$ is minimized when the input modes $\hat{\boldsymbol a} = [\begin{array}{cccc} \hat{a}_1 & \hat{a}_2 & \cdots & \hat{a}_n\end{array}]^T$ are in their vacuum states.  The resulting minimum von Neumann output entropy is $ng((1-\eta)K)$.}\\ 

Neither strong conjecture~1 nor its weak (single-use) form have been proven yet, but considerable evidence in support of their validity has been developed.  For example, strong conjecture~1 has been shown to be true when the input states are restricted to be Gaussian \cite{Eisert}.  It has also been proven that the vacuum state provides a local minimum for the output entropy \cite{gglms}.  Strong conjecture~1 has been shown to be true when  R\'{e}nyi entropy of integer order $\ge 2$ is employed in lieu of von Neumann entropy \cite{renyiproof}.  Similarly, conjecture~1 has been proven when Wehrl entropy---the continuous Boltzmann-Gibbs entropy of the Husimi probability function \cite{wehrl}---is used instead of von Neumann entropy \cite{renyiproof}.   Additional evidence in support of conjecture~1 can be found in \cite{gglms}.  

In proving the converse to the bosonic broadcast channel's capacity theorem we assumed the validity of the following conjecture.\\

\noindent{\bf {Strong Conjecture 2}} --- {\em Let the input modes $\hat{\boldsymbol a}$ be in their vacuum states, and let the von Neumann entropy of the input modes $\hat{\boldsymbol b}$  be $ng(K)$. Then, putting the $\hat{\boldsymbol b}$ modes in a product state of mean-photon-number $K$ thermal states minimizes the von Neumann entropy of the output modes $\hat{\boldsymbol c} = \sqrt{\eta}\,\hat{\boldsymbol a} + \sqrt{1-\eta}\,\hat{\boldsymbol b}$. The resulting minimum von Neumann output entropy is $ng((1-\eta)K)$.}\\

The weaker, single-use version of this conjecture is also of interest.\\

\noindent {\bf {Conjecture 2}} --- {\em Let the input mode $\hat{a}$ be in its vacuum state, and the let the von Neumann entropy of the input mode $\hat{b}$ be $g(K)$. Then the von Neumann entropy of the output mode $\hat{c} = \sqrt{\eta}\,\hat{a} + \sqrt{1-\eta}\,\hat{b}$ is minimized when $\hat{b}$ is in a thermal state with average photon number $K$. The resulting minimum von Neumann output entropy is $g((1-\eta)K)$.}\\

We have yet to develop proofs for either strong conjecture~2 or conjecture~2.  In the rest of this appendix we will present evidence that supports their validity.  Toward that end, we first show that strong conjecture~2 is true when Wehrl entropy is used instead of von Neumann entropy.  

\subsection{Strong Conjecture 2 for Wehrl Entropy}

\noindent {\bf {Strong Conjecture 2: Wehrl}} --- {\em Let the input modes $\hat{\boldsymbol a}$ be in their vacuum states, and let the Wehrl entropy of the input modes $\hat{\boldsymbol b}$  be $n(1+\ln(K+1))$. Then, putting the $\hat{\boldsymbol b}$ modes in a product state of mean-photon-number $K$ thermal states minimizes the Wehrl entropy of the output modes $\hat{\boldsymbol c} = \sqrt{\eta}\,\hat{\boldsymbol a} + \sqrt{1-\eta}\,\hat{\boldsymbol b}$. The resulting minimum Wehrl output entropy is $n(1+\ln{(K(1-\eta)+1)})$.}\\   

\noindent {\bf Proof} --- The Wehrl entropy for an $n$-mode density operator $\hat{\rho}$ is \begin{equation}
W({\hat \rho}) \equiv -\int{Q_{\hat{\rho}}(\boldsymbol{\mu})\ln{[\pi^n{Q_{\hat{\rho}}(\boldsymbol{\mu})}]}}d^{2n}\boldsymbol{\mu},
\label{eq:wehrldef}
\end{equation}
where $Q_{\hat{\rho}}(\boldsymbol{\mu}) \equiv \langle{\boldsymbol{\mu}}|{\hat \rho}|{\boldsymbol{\mu}}\rangle/{\pi^n}$, with $|\boldsymbol{\mu}\rangle$ a coherent state, is the Husimi distribution, i.e., the joint probability density function for multi-mode heterodyne detection. The Wehrl entropy provides a measurement of the state ${\hat \rho}$ in phase space and its minimum value is achieved on coherent states \cite{wehrl}. 

Our proof of strong conjecture~2 for Wehrl entropy relies on the antinormally-ordered characteristic function, $\chi_A^{\hat{\rho}_{\boldsymbol a}}({\boldsymbol{\zeta}})$, associated with the $n$-mode density operator ${\hat \rho}$, namely
\begin{equation}
\chi_A^{\hat{\rho}_{\boldsymbol a}}({\boldsymbol{\zeta}}) = {\rm tr}\left({\hat \rho}_{\boldsymbol a}e^{-{\boldsymbol{\zeta}}^\dagger{\hat {\boldsymbol a}}}e^{{\boldsymbol{\zeta}}^T{{\hat {\boldsymbol a}}^\dagger}}\right),
\label{eq:chi-def}
\end{equation}
where $\boldsymbol \zeta = [\begin{array}{cccc}\zeta_1 & \zeta_2 & \cdots & \zeta_n\end{array}]^T$ is a column vector of complex numbers, ${\boldsymbol \zeta}^\dagger = [\begin{array}{cccc}\zeta_1^* & \zeta_2^* & \cdots & \zeta_n^*\end{array}]$, and $\hat{\boldsymbol a}^\dagger = [\begin{array}{cccc} \hat{a}_1^\dagger & \hat{a}_2^\dagger & \cdots & \hat{a}_n^\dagger\end{array}]^T$. The antinormally-ordered characteristic function and the Husimi function are a 2$n$-D Fourier transform  pair:
\begin{eqnarray}
\chi_A^{\hat \rho}({\boldsymbol \zeta}) &=& \int{Q_{\hat \rho}({\boldsymbol \mu})e^{{\boldsymbol \mu}^\dagger{{\boldsymbol \zeta}}-{\boldsymbol \zeta}^\dagger{\boldsymbol \mu}}}d^{2n}{\boldsymbol \mu}, \\ 
Q_{\hat \rho}({\boldsymbol \mu}) &=& \frac{1}{\pi^{2n}}\int{\chi_A^{\hat \rho}({\boldsymbol \zeta})e^{-{\boldsymbol \mu}^\dagger{{\boldsymbol \zeta}}+{\boldsymbol \zeta}^\dagger{\boldsymbol \mu}}}d^{2n}{\boldsymbol \zeta}.
\end{eqnarray}

As the two $n$-use input modes $\hat{\boldsymbol a}$ and $\hat{\boldsymbol b}$ are in a product state, Eq.~(\ref{eq:chi-def}) implies that the output-state characteristic function is a product of the input-state characteristic functions with appropriately scaled arguments,
\begin{equation}
\chi_A^{\hat{\rho}_{\boldsymbol c}}({\boldsymbol \zeta}) = \chi_A^{\hat{\rho}_{\boldsymbol a}}(\sqrt{\eta}\,{\boldsymbol \zeta})\chi_A^{\hat{\rho}_{\boldsymbol b}}(\sqrt{1-\eta}\,{\boldsymbol \zeta}).
\label{eq:outputchi}
\end{equation}
From Eq.~(\ref{eq:outputchi}), the multiplication-convolution and scaling properties of Fourier-transforms pairs, and the fact that $\hat{\boldsymbol a}$ is in the $n$-mode vacuum state, we find that
\begin{eqnarray}
\lefteqn{Q_{\hat{\rho}_{\boldsymbol c}}({\boldsymbol \mu})} \nonumber \\
&=& \frac{1}{\eta^n}Q_{\hat{\rho}_{\boldsymbol a}}\left(\frac{{\boldsymbol \mu}}{\sqrt{\eta}}\right) \star \frac{1}{(1-\eta)^n}Q_{\hat{\rho}_{\boldsymbol b}}\left(\frac{{\boldsymbol \mu}}{\sqrt{1-\eta}}\right) \nonumber \\
&=& \frac{1}{(\pi\eta)^n}e^{-|{\boldsymbol \mu}|^2/\eta} \star \frac{1}{(1-\eta)^n}Q_{\hat{\rho}_{\boldsymbol b}}\left(\frac{{\boldsymbol \mu}}{\sqrt{1-\eta}}\right),
\label{eq:conv}
\end{eqnarray}
where $\star$ denotes convolution.

Suppose that the state of the input modes ${\hat {\boldsymbol b}}$ is a product of thermal states, each with mean photon number $K$, i.e., 
\begin{equation}
\hat{\rho}_{\boldsymbol b} = \left(\frac{1}{\pi{K}}\int{e^{-|\alpha|^2/K}|\alpha\rangle\langle\alpha|}d^2\alpha\right)^{\otimes n}.
\end{equation}
The Wehrl entropy for the $\hat{\boldsymbol b}$ modes is then 
\begin{equation}
W({\hat \rho_{\boldsymbol b}}) = n(1+\ln(K+1)),
\label{eq:thermalstateentropy}
\end{equation}
which satisfies the hypothesis of strong conjecture~2 for Wehrl entropy.  Using Eq.~(\ref{eq:conv}), we can now show that  the Husimi function and the Wehrl entropy for the state of the output modes $\hat {\boldsymbol c}$ are
\begin{eqnarray}
Q_{\hat{\rho}_{\boldsymbol c}}({\boldsymbol \mu}) &=& \frac{e^{-|{\boldsymbol \mu}|^2/(1+(1-\eta)K)}}{\left(\pi{(1+(1-\eta)K)}\right)^n} ,\\[.05in]
W({\hat{\rho}_{\boldsymbol c}}) &=& n(1+\ln(K(1-\eta)+1)), 
\end{eqnarray}
providing an upper-bound to the minimum Wehrl output entropy. 

To show that the expression in Eq.~(\ref{eq:thermalstateentropy}) is also a lower bound for the Wehrl output entropy, we use Theorem 6 of \cite{lieb}, which states that for two probability distributions $f({\boldsymbol \mu})$ and $h({\boldsymbol \mu})$ over $n$-dimensional complex vectors $\boldsymbol \mu$ we have
\begin{align}
&W((f \star h)({\boldsymbol \mu})) \ge \lambda{W(f({\boldsymbol \mu}))} + (1-\lambda)W(h({\boldsymbol \mu})) \nonumber \\
&\hspace*{.95in}- n\lambda\ln\lambda - n(1-\lambda)\ln(1-\lambda),
\label{eq:liebtheorem}
\end{align}
for all $0 \le \lambda \le 1$, where the Wehrl entropy of a probability distribution is found from Eq.~(\ref{eq:wehrldef}) by replacing $Q_{\hat{\rho}}({\boldsymbol \mu})$ with the given probability distribution. Choosing 
\begin{eqnarray}
f({\boldsymbol \mu}) &\equiv& \frac{1}{\eta^n}Q_{\hat{\rho}_{\boldsymbol a}}\left(\frac{{\boldsymbol \mu}}{\sqrt{\eta}}\right),  \\[.05in] 
h({\boldsymbol \mu}) &\equiv& \frac{1}{(1-\eta)^n}Q_{\hat{\rho}_{\boldsymbol b}}\left(\frac{{\boldsymbol \mu}}{\sqrt{1-\eta}}\right),
\label{eq:dists}
\end{eqnarray}
we get
\begin{eqnarray}
W({\hat{\rho}_{\boldsymbol c}}) &\ge& n\lambda(1+\ln\eta) \nonumber \\
&+& (1-\lambda)W\left(\frac{1}{(1-\eta)^n}Q_{\hat{\rho}_{\boldsymbol b}}\!\left(\frac{{\boldsymbol \mu}}{\sqrt{1-\eta}}\right)\right) \nonumber \\
&-& n\lambda\ln\lambda - n(1-\lambda)\ln(1-\lambda).
\label{eq:wehrl-LB-int}
\end{eqnarray}

The Wehrl entropy of a scaled distribution $(1/x)^nQ({\boldsymbol \mu}/\sqrt{x})$ is easily shown to satisfy
\begin{equation}
W\left(\frac{1}{x^n}Q\left(\frac{{\boldsymbol \mu}}{\sqrt{x}}\right)\right) = W\left(Q({\boldsymbol \mu})\right) + n\ln{x},
\label{eq:wehrl-scaling}
\end{equation}
for any $x > 0$. From Eqs.~(\ref{eq:wehrl-scaling}) and (\ref{eq:wehrl-LB-int}) we then obtain
\begin{eqnarray}
\lefteqn{W({\hat{\rho}_{\boldsymbol c}}) \ge n\lambda(1+\ln\eta) + (1-\lambda)\left(W({\hat{\rho}_{\boldsymbol b}})\right.} \nonumber \\
&+&\left.n\ln(1-\eta)\right) - n\lambda\ln\lambda - n(1-\lambda)\ln(1-\lambda) \nonumber \\
&=& n\lambda(1+\ln\eta) + n(1-\lambda)\left(1+\ln(K+1)\right. \nonumber \\
&+&\left.  \ln(1-\eta)\right) - n\lambda\ln\lambda - n(1-\lambda)\ln(1-\lambda) \nonumber \\
&=& n(1 + \ln(K(1-\eta)+1)).
\end{eqnarray}
The last equality used $\lambda = \eta/(\eta + (K+1)(1-\eta))$, which satisfies  $0\le \lambda \le 1$ for  all $\eta$ and $K$. Therefore the minimum Wehrl entropy of the output modes $\hat{\boldsymbol c}$ has the lower bound $n(1+\ln(K(1-\eta)+1))$.   
Because this lower bound coincides with the upper bound, derived earlier, we know that it is indeed the minimum Wehrl output entropy, and, moreover, that this minimum is achieved by a product thermal-state ${\hat{\rho}_{\boldsymbol b}}$ with mean photon number $K$ per mode.

\subsection{Strong Conjecture 2 for Gaussian-State Inputs}

In this section we prove that  strong conjectures~1 and 2 are equivalent when all inputs are restricted to be in Gaussian states.  Because strong conjecture~1 has been proven for Gaussian-state inputs \cite{Eisert}, this equivalence implies the truth of strong conjecture~2 for such inputs.  

With no loss of generality we shall restrict our attention to zero-mean Gaussian states.  A zero-mean $n$-mode Gaussian state is completely characterized by its $2n \times 2n$ correlation matrix 
\begin{equation}
{\bf R}_{\boldsymbol a} = \left \langle \begin{bmatrix} \boldsymbol{\hat{a}} \\ \boldsymbol{\hat{a}^{\dagger}} \end{bmatrix} \begin{bmatrix} (\boldsymbol{\hat{a}^{\dagger}} )^{T} & \boldsymbol{\hat{a}}^{T} \end{bmatrix} \right \rangle = \begin{bmatrix}  \langle \boldsymbol{\hat{a}^{\dagger}} \boldsymbol{\hat{a}}^{T} \rangle + {\bf I}_{n} & \langle \boldsymbol{\hat{a}} \boldsymbol{\hat{a}}^{T} \rangle \\ \langle \boldsymbol{\hat{a}} \boldsymbol{\hat{a}}^{T} \rangle^{*} & \langle \boldsymbol{\hat{a}^{\dagger}} \boldsymbol{\hat{a}}^{T} \rangle \end{bmatrix}\,,
\label{corrmat}
\end{equation}
where ${\bf I}_{n}$ is the $n \times n$ identity matrix and $*$ denotes component-wise complex conjugation.  Of particular importance, for what will follow, is the symplectic diagonalization of ${\bf R}_{\boldsymbol a}$ and the consequences thereof.  

Let the $n$ modes represented by $\boldsymbol{\hat{a}}$ be in a zero-mean Gaussian state with  correlation matrix ${\bf R}_{\boldsymbol a}$. We will show that there exists a $2n\times 2n$ complex-valued symplectic matrix ${\bf S}$  such that 
\begin{equation}
{\bf R}_{\boldsymbol a} = {\bf S} {\bf \Lambda}  {\bf S}^{\dagger}\,, 
\label{symp1} 
\end{equation}
where ${\bf S}^\dagger$ is the conjugate transpose of ${\bf S}$,
\begin{equation} 
{\bf S}^{\dagger} {\bf Q} {\bf S} = {\bf S} {\bf Q} {\bf S}^{\dagger} = {\bf Q},
\label{symplectcondx}
\end{equation}
and 
\begin{equation}
{\bf \Lambda} = 
{\rm diag} [\begin{array}{cccccc}\lambda_{1}+1  & \cdots & \lambda_{n}+1 & \lambda_{1} & \cdots & \lambda_{n}\end{array}].  
\label{symp2}
\end{equation} 
Equation~\eqref{symplectcondx}, with 
\begin{equation}
{\bf Q} = \begin{bmatrix} {\bf I}_{n} & 0 \\ 0 & - {\bf I}_{n} \end{bmatrix},
\end{equation} 
is the condition that makes ${\bf S}$ symplectic.  The $\{\lambda_i\}$ are the symplectic eigenvalues of ${\bf R}_{\boldsymbol a}$, which are all non-negative because ${\bf R}_{\boldsymbol a}$ is positive semidefinite.  

To establish the preceding symplectic diagonalization of ${\bf R}_{\boldsymbol a}$, we use 
Williamson's symplectic decomposition theorem on the symmetrized (real-valued) correlation matrix for the quadratures, $\hat{\boldsymbol a}_{1} \equiv {\rm Re}(\hat{\boldsymbol a})$ and  $\hat{\boldsymbol a}_{2} \equiv {\rm Im}(\hat{\boldsymbol a})$  \cite{Gosson}. Equations~\eqref{symp1}--\eqref{symp2} are then obtained by converting this quadrature correlation-matrix decomposition into the annihilation operator correlation matrix via the linear transformation
\begin{equation}
{\bf U} = \begin{bmatrix}{\bf I}_{n} & i {\bf I}_{n} \\ {\bf I}_{n} & -i {\bf I}_{n} \end{bmatrix}\,.
\end{equation}

The value of the symplectic decomposition lies in establishing a linear relationship between the modes ${\hat{\boldsymbol a}}$, which are in a given zero-mean Gaussian state, to a new set of modes that are in independent thermal states whose average photon numbers are given by the symplectic eigenvalues.
In particular, for $\boldsymbol{\hat{a}}$ in an arbitrary $n$-mode zero-mean Gaussian state with correlation matrix ${\bf R}_{\boldsymbol a}$, we have that
\begin{equation}
\begin{bmatrix} {\hat{\boldsymbol d}} \\ {\hat{\boldsymbol d}^{\dagger}} \end{bmatrix} = {\bf S}^{-1} \begin{bmatrix}  {\hat{\boldsymbol a}} \\ {\hat{\boldsymbol a}^{\dagger}} \end{bmatrix}
\label{symptrans}
\end{equation}
accomplishes this transformation, where ${\bf S}^{-1} = {\bf Q}{\bf S}^{\dagger} {\bf Q}$.  The $n$ modes represented by $\hat{\boldsymbol d}$ are in a zero-mean Gaussian state with a correlation matrix that is easily found to be 
\begin{equation}
{\bf R}_{\boldsymbol d} = {\bf \Lambda}\,.
\end{equation}
Thus, $\hat{d}_{i}$ has average photon number $\langle \hat{d}_{i}^{\dagger} \hat{d}_{i} \rangle = \lambda_{i}$, for $1\le i \le n$.  Furthermore, the $\hat{d}_i$ modes are all uncorrelated---because ${\bf \Lambda}$ is diagonal---so that each mode can be represented as an isotropic Gaussian mixture of coherent states, and the joint state is the tensor product of $n$ such states.

The symplectic transformation in \eqref{symp1} is canonical, i.e., it preserves the commutation relations.  Thus it can be implemented with a unitary operator $\hat{U}$, satisfying $\hat{U} \hat{U}^{\dagger}= \hat{U}^{\dagger} \hat{U} = \hat{I}$ \cite{Yuen:TwoPhoton,MaRhodes}.
From this we see that the von Neumann entropy of the Gaussian-state ${\hat{\boldsymbol a}}$ modes is identical to that of the thermal-state ${\hat{\boldsymbol d}}$ modes from \eqref{symptrans}.  We are now ready to address the central concern of this section, namely showing that strong conjecture~2 is true when the input states are restricted to be Gaussian.  

\begin{thm}
Strong conjecture~1 and strong conjecture~2 are equivalent when the input fields are restricted to be in  Gaussian states.
\end{thm}

\begin{proof}
Consider the the vector input-output relation
\begin{equation}
\begin{bmatrix} {\hat{\boldsymbol c}} \\ {\hat{\boldsymbol c}}^{\dagger} \end{bmatrix} =  \sqrt{\eta } \begin{bmatrix} {\hat{\boldsymbol a}} \\ {\hat{\boldsymbol a}}^{\dagger} \end{bmatrix} + \sqrt{1-\eta} \begin{bmatrix} {\hat{\boldsymbol b}} \\ {\hat{\boldsymbol b}}^{\dagger} \end{bmatrix}\,,
\label{BS:vector}
\end{equation}
with the $\hat{\boldsymbol a}$ and $\hat{\boldsymbol b}$ modes being in independent quantum states that are zero-mean Gaussians.  

First let us use the truth of strong conjecture~1 to show that strong conjecture~2 is also true.  Under the premise of strong conjecture~2, we take the $\hat{\boldsymbol a}$ modes to be in their vacuum states,
and the $\hat{\boldsymbol b}$ modes to be in a zero-mean Gaussian state with correlation matrix ${\bf R}_{\boldsymbol b}$ and von Neumann entropy $ng(K)$.  (No loss in generality ensues from the restriction that the $\hat{\boldsymbol b}$ modes be in a zero-mean state, because von Neumann entropy is invariant to state displacement.)   Because the inputs are in Gaussian states, minimizing the von Neumann entropy of the output modes $\hat{\boldsymbol c}$ reduces to finding the correlation matrix ${\bf R}_{\boldsymbol b}$ that minimizes this entropy.   Let ${\bf R}_{\boldsymbol b} = {\bf S} {\bf \Lambda} {\bf S}^{\dagger}$ be the symplectic diagonalization of ${\bf R}_{\boldsymbol b}$.   We can then express the input modes $\hat{\boldsymbol b}$ as a symplectic transformation on another set of $n$-modes, $\hat{\boldsymbol d}$,
\begin{equation}
\begin{bmatrix} {\hat{\boldsymbol b}} \\ {\hat{\boldsymbol b}}^{\dagger} \end{bmatrix} = {\bf S} \begin{bmatrix} {\hat{\boldsymbol d}} \\ {\hat{\boldsymbol d}}^{\dagger} \end{bmatrix}\,,
\label{sympdecomp:Forward}
\end{equation}
whose correlation matrix is ${\bf R}_{\boldsymbol d} = {\bf \Lambda}$.  Furthermore, we have that 
\begin{equation}
S(\hat{\rho}_{\boldsymbol d}) = \sum_{i=1}^{n} S(\hat{\rho}_{d_i}) = \sum_{i=1}^{n} g(\lambda_{i}) = S(\hat{\rho}_{\boldsymbol b}) = ng(K).
\end{equation}
\begin{figure}[t]
\centering
\includegraphics[width=3in]{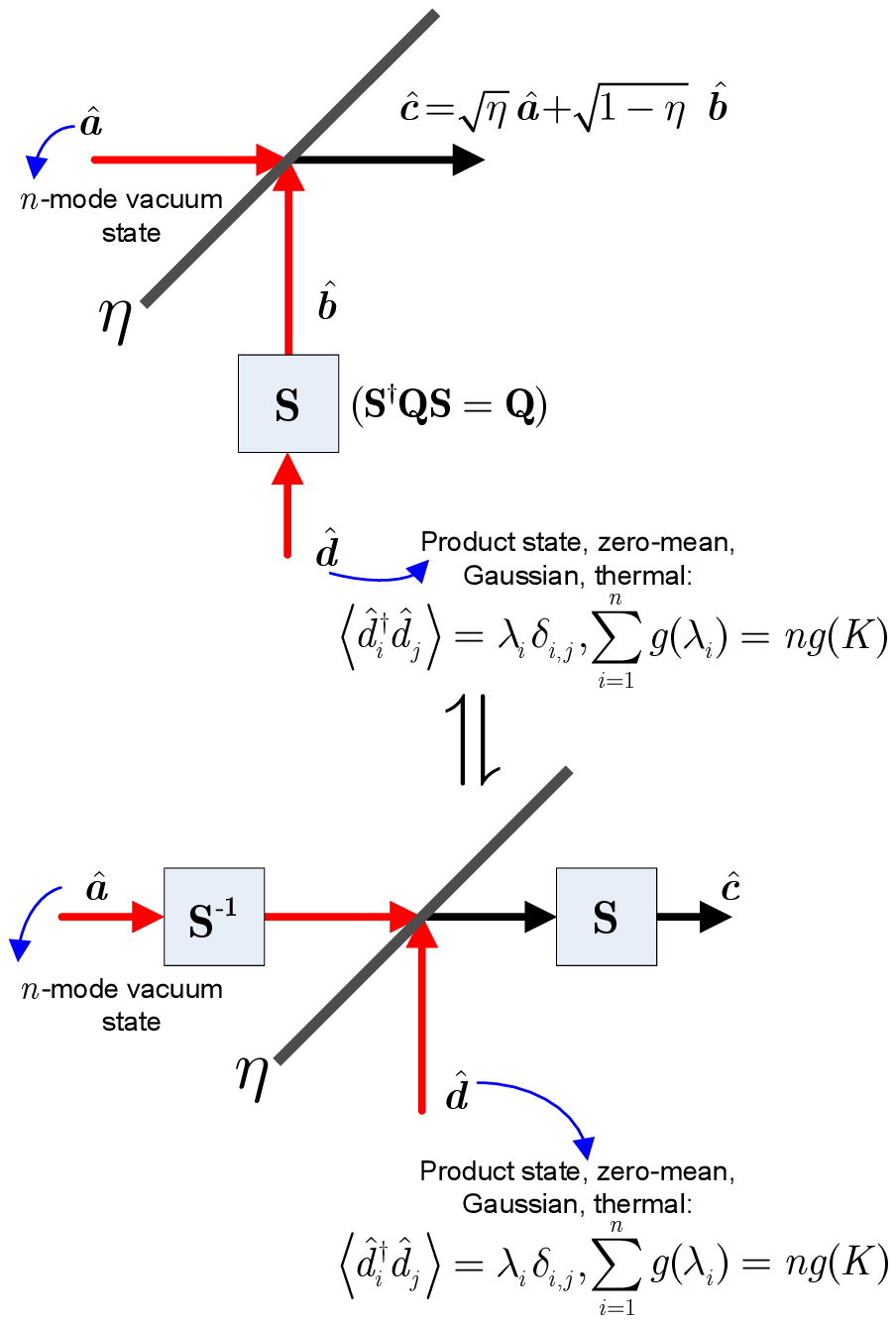}
\caption{(Color online) Schematic of the beam splitter channel with zero-mean Gaussian-state inputs, and the equivalent channel after transformations that preserve von Neumann entropies.}
\label{Conj1Fig}
\end{figure}

Substituting Eq.~\eqref{sympdecomp:Forward} into \eqref{BS:vector} and using some linear algebra, we get
\begin{equation}
\begin{bmatrix} {\hat{\boldsymbol c}} \\ {\hat{\boldsymbol c}}^{\dagger} \end{bmatrix} ={\bf S} \left (\sqrt{\eta}\, {\bf S}^{-1} \begin{bmatrix} {\hat{\boldsymbol a}} \\ {\hat{\boldsymbol a}}^{\dagger} \end{bmatrix} + \sqrt{1-\eta} \begin{bmatrix}{\hat{\boldsymbol d}} \\ {\hat{\boldsymbol d}}^{\dagger} \end{bmatrix}\right)\,.
\end{equation}
A schematic corresponding to this equation is shown in the bottom panel of Fig.~\ref{Conj1Fig}. In particular, the beam splitter channel governed by Eq.~\eqref{BS:vector} and the Gaussian states we have assumed for $\hat{\boldsymbol a}$ and $\hat{\boldsymbol b}$ are equivalent to what we have shown in the top panel of Fig.~\ref{Conj1Fig}.   We know that symplectic transformations do not affect von Neumann entropy.  Thus minimizing the von Neumann entropy of the $\hat{\boldsymbol c}$ modes by choice of the correlation matrix ${\bf R}_{\boldsymbol b}$ in the top panel of Fig.~\ref{Conj1Fig} is equivalent to minimizing this output entropy by choice of the $2n \times 2n$ symplectic matrix ${\bf S}$ and the symplectic eigenvalues $\{\lambda_{i}\geq 0 : 1\le i \le n\}$, subject to the constraint that $\sum_{i=1}^n g(\lambda_{i}) = ng(K)$, in the lower panel of that figure. 

	Suppose that we have a set of symplectic eigenvalues that satisfy the constraint. Then, via strong conjecture~1, the von Neumann entropy the $\hat{\boldsymbol c}$ modes is minimized when the $\hat{\boldsymbol a}$ modes in the lower panel of Fig.~\ref{Conj1Fig} are in their vacuum states.  However, because the $\boldsymbol{\hat{a}}$ modes are already in this state, an optimizing symplectic transformation ${\bf S}^{-1}$ is the identity matrix ${\bf I}_{2n}$.  This result is independent of the particular values of the $\{\lambda_{i}\}$, but the entropy of the $\hat{\boldsymbol c}$ modes still depends on our choice of symplectic eigenvalues.  In particular, when the $\hat{\boldsymbol a}$ modes are in their vacuum states and ${\bf S}^{-1} = {\bf I}_{2n}$, the von Neumann entropy of the $\hat{\boldsymbol c}$ modes is 
\begin{equation}
S(\hat{\rho}_{\boldsymbol c}) = \sum_{i=1}^{n} g(\sqrt{1-\eta}\, \lambda_{i})\,.
\end{equation}
To minimize this output entropy, by choice of the $\{\lambda_{i}\}$ we employ a Lagrange multiplier approach:
\begin{multline}
\min_{\{\lambda_{i} \geq 0 : \sum_{i=1}^n g(\lambda_{i}) = ng(K)\}}{\sum_{i=1}^{n} g(\sqrt{1-\eta}\, \lambda_{i})} = \\  \min_{\lambda_{i} \geq 0, \xi} {\left \{ \sum_{i=1}^{n} g(\sqrt{1-\eta}\,\lambda_{i})-\xi \left(\sum_{i=1}^{n} g(\lambda_{i}) - ng(K)\right)\right \}}.
\label{lagrangemult}
\end{multline}
Differentiating Eq.~\eqref{lagrangemult} with respect to the $\{\lambda_{i}\}$ and $\xi$ yields
\begin{align}
&\xi = \frac{\sqrt{1-\eta}\, g'(\sqrt{1-\eta} \, \lambda_{i})}{g'(\lambda_{i})}, \quad\mbox{for  $1\le i \le n$} \\
&\sum_{i=1}^{n} g(\lambda_{i}) = ng(K)\,,
\end{align}
which implies that choosing $\lambda_{i} = K$, for $1\le i \le n$, minimizes the output entropy subject to the constraint \cite{footnote4}. The minimum output entropy is then $n g(\sqrt{1-\eta} K)$, and it is achieved when the $n$-mode Gaussian input state is an $n$-mode thermal product state with $\langle \hat{b}_{i}^{\dagger} \hat{b}_{i} \rangle = K$ for  $1 \le i \le n$.  This completes the demonstration that strong conjecture~1 implies strong conjecture~2 for Gaussian-state inputs, and because strong conjecture~1 is known to be true for Gaussian-state inputs, we have proven that strong conjecture~2 is also true for such inputs.  To complete the proof of Theorem~1, we must still show that strong conjecture~2 implies strong conjecture~1 when the input states are Gaussian.
 
Assume that strong conjecture~2 is true, and let the input modes $\hat{\boldsymbol b}$ be in a product state of $n$ zero-mean thermal states each with von Neumann entropy $g(K)$, as shown in the top panel of Fig.~\ref{Conj2Fig}. With no loss of generality we can take the input modes $\hat{\boldsymbol a}$ to be in a zero-mean pure Gaussian state, i.e., $\boldsymbol{\hat{a}}$ is in an $n$-mode vacuum or squeezed-vacuum state.  Performing the symplectic diagonalization ${\bf R}_{\boldsymbol a} = {\bf S} {\bf \Lambda} {\bf S}^{\dagger}$, we write 
\begin{equation}
\begin{bmatrix} {\hat{\boldsymbol a}} \\ {\hat{\boldsymbol a}}^{\dagger} \end{bmatrix} = {\bf S} \begin{bmatrix} {\hat{\boldsymbol d}} \\ {\hat{\boldsymbol d}}^{\dagger} \end{bmatrix}\,,
\label{sympdecomp:Backward}
\end{equation}
where ${\bf R}_{\boldsymbol d} = {\bf \Lambda}$.  
Because this transformation preserves von Neumann entropy, we know that $\hat{\rho}_{\boldsymbol d}$ must be a zero-mean pure Gaussian state with no phase-sensitive correlation. The only such state is the $n$-mode vacuum state. 
We can then perform similar algebraic manipulations to the beam splitter relation in Eq.~\eqref{BS:vector} to get,
\begin{equation}
\begin{bmatrix} {\hat{\boldsymbol c}} \\ {\hat{\boldsymbol c}}^{\dagger} \end{bmatrix} = {\bf S} \left (\sqrt{\eta} \begin{bmatrix} {\hat{\boldsymbol d}} \\ {\hat{\boldsymbol d}}^{\dagger} \end{bmatrix} + \sqrt{1-\eta}\, {\bf S}^{-1}  \begin{bmatrix} {\hat{\boldsymbol b}} \\ {\hat{\boldsymbol b}}^{\dagger} \end{bmatrix}\right)\,,
\end{equation}
as shown in the lower panel in Figure~\ref{Conj2Fig}.  
\begin{figure}[t]
\centering
\includegraphics[width=3in]{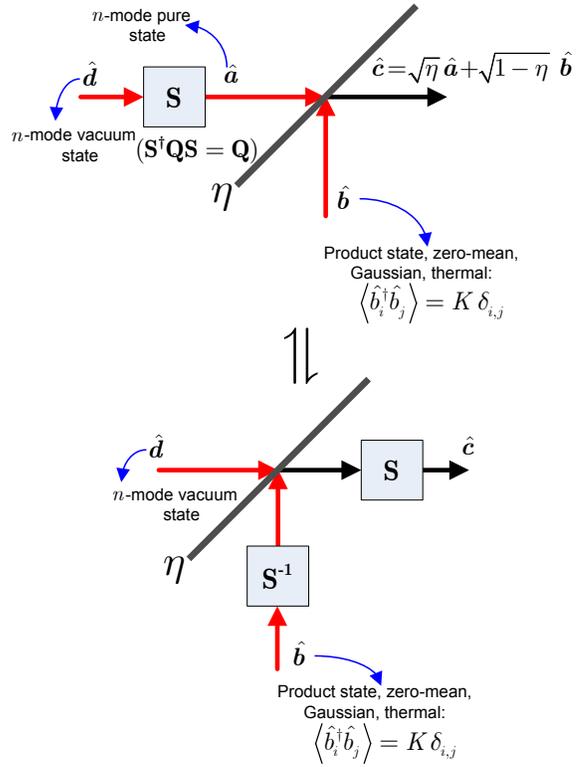}
\caption{(Color online) Schematic of the beam splitter channel with zero-mean Gaussian-state inputs, and the equivalent channel after transformations that preserve von Neumann entropy.}
\label{Conj2Fig}
\end{figure}

Minimizing the von Neumann entropy after the symplectic transformation at the output port in the lower panel of Fig.~\ref{Conj2Fig} is equivalent to minimizing the entropy before that transformation. Thus our objective is to determine the $2n \times 2n$ symplectic matrix ${\bf S}^{-1}$ that minimizes the von Neumann entropy before the output-port symplectic transformation.  Because the $\hat{\boldsymbol a}$ modes are in their vacuum states and the modes applied to the beam splitter's other input port have von Neumann entropy $ng(K)$, strong conjecture~2 tells us that the latter input should be in an $n$-mode thermal product state, with average photon number $K$ per mode, to achieve the minimum output entropy. But ${\hat{\boldsymbol b}}$ is already in this state, so an optimizing symplectic transformation is therefore the identity, ${\bf S}^{-1} = {\bf I}_{2n}.$
This allows us to conclude that putting the $\hat{\boldsymbol a}$ modes in their vacuum states  minimizes the entropy of the $\hat{\boldsymbol c}$ modes when the $\hat{\boldsymbol b}$ modes are  in an $n$-mode product of thermal states each with average photon number $K$, thus demonstrating that strong conjecture~2 implies strong conjecture~1 when the input states are Gaussian.
\end{proof}

\section{A property of $g(x)$}

For the converse proof given in Sec.~IV, we need to show that for non-negative real numbers $\{x_k  : 1\le k \le n\}$, and $0 \le \eta \le 1$, if $x_0$ is defined by
\begin{equation}
\sum_{k=1}^n\frac{g(x_k)}{n} = g(x_0),
\end{equation}
then 
\begin{equation}
\sum_{k=1}^n\frac{g(\eta{x_k})}{n} \ge g(\eta{x_0}),
\end{equation}
where $g(x) \equiv (1+x)\log(1+x) - x\log(x)$. 

\begin{figure}
\begin{center}
\includegraphics[width=3in]{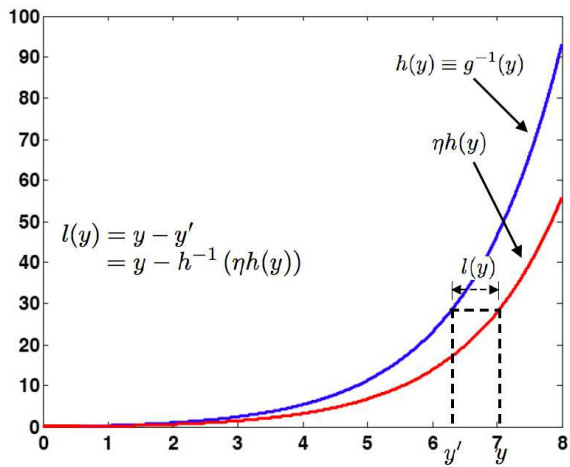}
\end{center}
\caption{(Color online) Plots of the 1-to-1 function $g(x)$ and its inverse $h(y)$, showing $y^\prime \equiv h^{-1}\left(\eta{h(y)}\right) = g\left(\eta{g^{-1}(y)}\right)$ and $l(y) = y - y^\prime$.}
\label{fig:hy}
\end{figure}

Because $g(x)$ is a 1-to-1 function, it has an inverse function $h(y) \equiv g^{-1}(y)$, such that if $y = g(x)$ then $x = h(y)$. Let $y_k = g(x_k)$, for $1\le k \le n$. For every $y \ge 0$, define $y^\prime = h^{-1}\left(\eta{h(y)}\right) = g\left(\eta{g^{-1}(y)}\right)$ and define $l(y) = y - y^\prime$, as shown in Fig.~\ref{fig:hy}.   Using this notation what we are trying to prove becomes the following.  Given that
\begin{equation}
y_0 = \frac1n\sum_{k=1}^ny_k,
\label{th2eq1}
\end{equation}
show that
\begin{equation}
\frac1n\sum_{k=1}^ny_k^\prime \ge y_0^\prime.
\end{equation}

By straightforward differentiation, we can show that ${d^2}l(y)/dy^2 \le 0$ which implies that
\begin{eqnarray}
y_0 - y_0^\prime \ge  \frac1n\sum_{k=1}^ny_k -  \frac1n\sum_{k=1}^ny_k^\prime,\\ 
\nonumber
\end{eqnarray}
from the definition of $l(y)$.  
Using Eq.~(\ref{th2eq1}) we then get
\begin{equation}
\frac1n\sum_{k=1}^ny_k^\prime \ge y_0^\prime,
\end{equation}
which completes the proof.


\begin{thebibliography}{99}
\bibitem{footnote1}${\cal A}^n$, ${\cal B}^n$, and ${\cal C}^n$ are the $n$ channel use alphabets of Alice, Bob, and Charlie, with respective sizes $|{\cal A}^n|$, $|{\cal B}^n|$, and $|{\cal C}^n|$.
\bibitem{footnote3}Holevo's bound \cite{holevo}: Let $X$ be the input alphabet for a channel,  $\left\{p_i, {\hat \rho}_i\right\}$ be the priors and modulating states, $\left\{\Pi_j\right\}$ be a POVM, and $Y$ the resulting output (classical) alphabet. The Shannon mutual information $I(X;Y)$ cannot exceed the Holevo information $\chi(p_i,{\hat \rho}_i)$.
\bibitem{footnote4}This solution is unique because the derivative, $g'(x)$, is an invertible function.
\end{thebibliography}
\end{document}